\documentclass[12pt,a4paper,twoside]{article}
\usepackage{epsfig}
\usepackage{rotating}
\usepackage{cite}
\long\def\ifundef#1#2#3{\expandafter\ifx\csname
  #1\endcsname\relax#2\else#3\fi}

\font\sfhuge= cmss24  at 20truept
\font\sfLARGE= cmss14  at 14truept
  at 12truept
\font\sfmed= cmss10  at 10truept
   at  8truept
\ifundef{ensuremath}{\newcommand{\ensuremath}[1] {\relax\ifmmode {#1}\else {$#1$} \fi}}{}
\ifundef{mathcal}{\newcommand{\mathcal}[1]{{\cal #1}}}{}
\ifundef{mathrm}{\newcommand{\mathrm}[1]{\rm #1}}{}
\ifundef{resizebox}{}{}
\ifundef{includegraphics}{\newcommand{\includegraphics}[1] 
{$#1$ 
\typeout{PostScript File #1 not included!}
}}{}

\newcommand{\epem}   {\ensuremath{\mathrm{e^+e^-}}}

\newcommand{\as}     {\ensuremath{\alpha_s}}

\newcommand{\asmz}   {\ensuremath{\alpha_s(M_{\mathrm{Z^0}})}}
\newcommand{\azero}     {\ensuremath{\alpha_0}}
\newcommand{\azeromuI}  {\ensuremath{\alpha_0(\mu_I})}
\newcommand{\azerotwo}  {\ensuremath{\alpha_0({\mathrm{2~GeV}})}}
\newcommand{\muI}  {\ensuremath{\mu_I}}

\newcommand{\oaa}    {\ensuremath{\mathcal{O}(\alpha_s^2)}}

\newcommand{\mz}     {\ensuremath{M_{\mathrm{Z^0}}}}

\newcommand{\bt}     {\ensuremath{B_T}}
\newcommand{\bw}     {\ensuremath{B_W}}

\newcommand{\chisq}  {\ensuremath{\chi^2}}
\newcommand{\chisqd} {\ensuremath{\chi^2/\mathrm{d.o.f.}}}
\newcommand{\xmu}    {\ensuremath{x_{\mu}}}

\ifundef{pt}{ }
            { }

\newcommand{\rs}     {\ensuremath{\sqrt{s}}}

\newcommand{\bm}[1]  {\mbox{\boldmath\ensuremath{#1}}}

\newcommand{\s}      {\textstyle}
\newcommand{\lt}     {\ensuremath{<}}
\newcounter{hours}
\newcounter{minutes}

\newcommand{\Printtime}{%
  \setcounter{hours}{\time/60}%
  \setcounter{minutes}{\time-\value{hours}*60}%
  \ifthenelse{\value{hours}<10}{0}{}\thehours:%
  \ifthenelse{\value{minutes}<10}{0}{}\theminutes}

\begin{document}
\addtolength{\textheight}{-68mm}
\bibliographystyle{utphys}
%
%  Title Page (PITHA cover)
\begin{titlepage}
\pagestyle{empty}

\vspace*{-10mm}
\vbox to 245mm{

\hbox to \textwidth{ \hsize=\textwidth
\vbox{
\hbox{
\epsfig{file=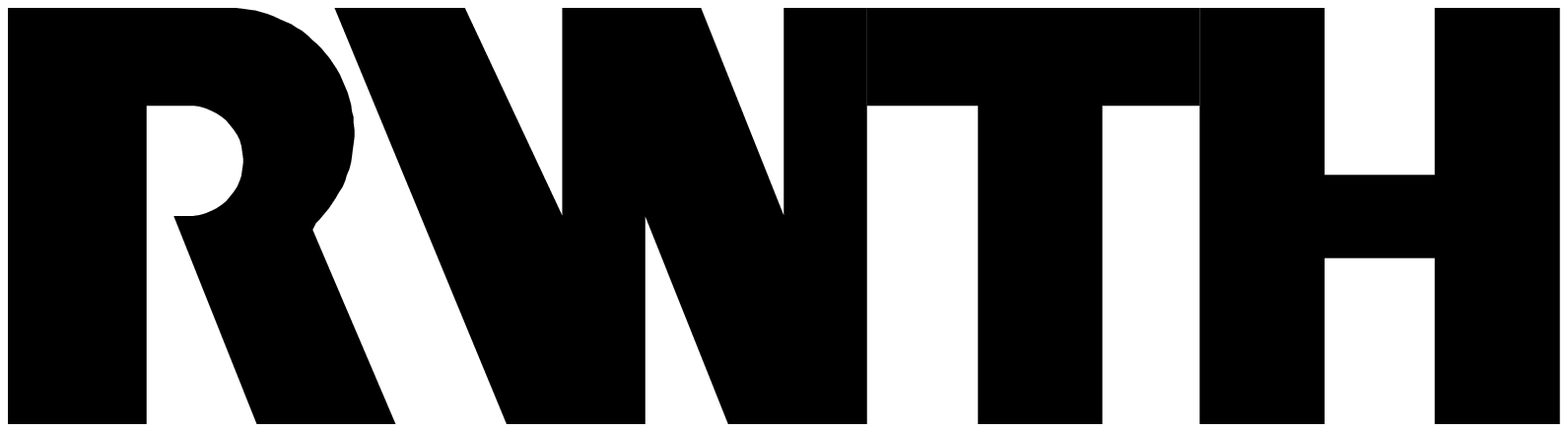,height=20mm}
}
}
\vbox{
{
\hbox{\sfmed RHEINISCH-\hss}\vspace*{-.73mm}
\hbox{\sfmed WESTF\"ALISCHE-\hss}\vspace*{-.73mm}
\hbox{\sfmed TECHNISCHE-\hss}\vspace*{-.73mm}
\hbox{\sfmed HOCHSCHULE-\hss}\vspace*{-.73mm}
\hbox{\sfmed AACHEN\hss}\vspace*{+0.200mm}
}
}
\vbox{ \hsize=58mm 
{
\hspace*{0pt\hfill}\hbox{\sfLARGE\hspace*{0pt\hfill}        PITHA 99/21\hss}\vspace*{-2mm}
\hspace*{0pt\hfill}\hbox{        \hspace*{0pt\hfill} \rule{45mm}{1.0mm}\hss}
\hspace*{0pt\hfill}\hbox{\sfLARGE\hspace*{0pt\hfill}       Juni 1999\hss}\vspace*{2.3mm}
}
}
}

\vspace*{5cm}

\begin{center}
{\huge\bf  Tests of Power Corrections to \\[2.3mm]  Event Shape Distributions   \\[2.7mm] from  e\ensuremath{^+}e\ensuremath{^-} Annihilation }

\end{center}
\vspace*{2cm}
\begin{center}
\Large
P.A.~Movilla~Fern\'andez, O.~Biebel, S.~Bethke \\
\bigskip 
\bigskip
III. Physikalisches Institut, Technische Hochschule Aachen\\
D-52056 Aachen, Germany
\end{center}

\vspace*{0pt\vfill}
\vfill

\vspace*{-5mm}
\noindent
\hspace*{-5mm}
\hbox {
\rule{\textwidth}{0.3mm}
}

\vspace*{3mm}
\noindent
\begin{minipage}{\textwidth}

\vbox {\vsize=60mm
\hbox to \textwidth{\hsize=\wd0
\hbox {\hspace*{-5mm}

\vbox{ 
\hbox to \textwidth{\hss\sfhuge PHYSIKALISCHE INSTITUTE\hss }\vspace*{2.0mm}
\hbox to \textwidth{\hss\sfhuge      RWTH AACHEN\hss }\vspace*{2.0mm}
%\hbox to \textwidth{\hss\sfhuge    Physikzentrum\hss }\vspace*{2.0mm}
\hbox to \textwidth{\hss\sfhuge D-52056 AACHEN, GERMANY\hss}
}

}
}
}

\end{minipage}
}
%\newpage\ 
%\newpage
\end{titlepage}

%
%  Title Page:
\begin{titlepage}

\pagestyle{empty}
%\vspace*{-10mm}
%\vbox{ \hsize=58mm
%{
%  \hbox{ EPS-HEP 99 \hss}
%  \hbox{ Abstract \# 1\_113\hss } 
%  \hbox{ Parallel session 1 \hss}
%  \hbox{ Plenary session 1 \hss}
%}
%}
%\vspace*{-19.6mm}
\hbox to \textwidth{ \hsize=\textwidth
\hspace*{0pt\hfill} 
\vbox{ \hsize=58mm
{
  \hbox{ PITHA 99/21\hss}
  \hbox{ Juni 11, 1999\hss } 
}
}
}

\bigskip\bigskip\bigskip\bigskip\bigskip\bigskip\bigskip\bigskip%\bigskip
\begin{center}
  {\Huge\bf
    Tests of Power Corrections to \\[2.3mm] 
    Event Shape Distributions  \\[2.7mm]
    from  e\ensuremath{^+}e\ensuremath{^-} Annihilation
    }
\end{center}
\bigskip\bigskip
\begin{center}
  P.A.~Movilla~Fern\'andez$^{(1)}$,
  O.~Biebel$^{(1)}$, 
  S.~Bethke$^{(1)}$ \\
\end{center}
\bigskip
%  Abstract:
\begin{abstract}
\noindent 
A study of differential event shape distributions using e$^+$e$^-$
data at centre-of-mass energies of \rs\ = 35 to 183~GeV is presented. We
investigated non-perturbative power corrections for the thrust,
C-parameter, total and wide jet broadening observables.  We observe a
good description of the distributions by the combined resummed QCD
calculations plus power corrections from the dispersive approach.  The
single non-perturbative parameter \azero\ is measured to be
\[
 \azerotwo = 0.502 \pm 0.013 \mathrm{(stat.)}
              \ ^{+0.046} 
                _{-0.032} \mathrm{(exp. \; syst.)}
              \ ^{+0.074} 
                _{-0.053} \mathrm{(theo. \; syst.)}
\]
and is found to be universal for the observables studied within the given
systematic uncertainties.
Using revised calculations of the power corrections for the jet
broadening variables, improved consistency of the individual fit results is
obtained.  Agreement is also found with results extracted from the
mean values of event shape distributions.
\end{abstract}
\vspace*{0pt\vfill}
\vfill
\bigskip\bigskip\bigskip\bigskip
{
\small
\noindent
$^{(1)}$ 
\begin{minipage}[t]{155mm} 
III. Physikalisches Institut der RWTH Aachen,
D-52056 Aachen, Germany \\
contact e-mail: Pedro.Fernandez@Physik.RWTH-Aachen.DE \\
\hspace*{25mm}  Otmar.Biebel@Physik.RWTH-Aachen.DE
\end{minipage}\\
}
%\newpage\ 
%\newpage
\end{titlepage}

\section{ Introduction }
\setcounter{page}{1}

The study of final states in electron-positron annihilation led to
the development of numerous observables as powerful tools
for the analysis of hadronic events. Improving theoretical
predictions for these observables in the last years as well as the
large amount of data collected by various experiments e.g. at the
PETRA, PEP, LEP and SLC accelerators from \rs~= 12 to 189~GeV provide
precise quantitative tests of Quantum Chromodynamics (QCD). The key
feature of perturbative QCD, the running of the strong coupling
constant \as, has meanwhile become an indubitable experimental
fact from \epem-data (see~\cite{bib-world-alphas-sb} and
references therein), in particular due to investigation of the 
energy dependence of event shape data.

Precision tests of perturbative QCD from hadronic event shapes require
a solid understanding of non-perturbative effects.  In the context of
hadronisation, non-perturbative effects are usually estimated by
phenomenological hadronisation models available from Monte Carlo event
generators.  Alternatively, analytical approaches were pursued in the
past years in order to deduce as much information as possible about
hadronisation from the perturbative theory, e.g. the renormalon
inspired ansatz~\cite{bib-renormalons} or the dispersive
approach\cite{bib-webber}.

Phenomenological considerations lead to the expectation that
hadronisation contributions to event shape observables evolve like
reciprocal powers of the hard interaction scale $\rs$.  The
investigation of power corrections therefore substantially benefits
from \epem data at the lower energy region as marked by PETRA/PEP
experiments at \rs~= 12 to 47 GeV.  Our re-analysis of \epem data
collected by the JADE detector~\cite{bib-newJADE,bib-newJADE2}, one of
the former PETRA experiments, allowed studies of the energy evolution
of QCD at {\em combined} lower and higher energies in terms of observables
frequently used in the analysis of events at the highest
centre-of-mass energies available at LEP.
Further efforts were undertaken to make event shapes available at
hadronic centre-of-mass energies below the Z pole, e.g. the analysis
of radiative events recorded at LEP-1 as performed by several LEP
experiments~\cite{bib-FSR}.

In the present paper we test power corrections to the differential
distributions of event shape observables measured from re-analysed
JADE data at 35 to 44~GeV and from data of the LEP/SLC
experiments at 91 to 183~GeV.  We focussed on two-loop
predictions~\cite{bib-Milan-factor,bib-new-broadening} basing on the
dispersive approach. For the jet broadening measures, progress was
made in the understanding of non-perturbative corrections to the
distributions \cite{bib-new-broadening}, in particular the
interdependence between perturbative and non-perturbative
contributions.  We also test a basic concept of the power corrections,
the universality of the zeroth moment of the effective coupling \azero\ 
below a given infrared energy scale \muI.

Section~\ref{sec-data} starts with an overview of the observables and
the data used in the analysis. After some annotations to the
parametrisation of the power corrections in Section~\ref{sec-powcor},
we present fit results basing on combined power corrections plus
perturbative QCD in Section~\ref{sec-tests} and draw conclusions from
our results in Section~\ref{sec-conclusions}.

\section{ Event shape data }
\label{sec-data}
For our studies, we considered the event shape distributions of
thrust, the $C$-parameter, the total and wide jet broadening, \bt\ and
\bw. For convenience, we list the definitions of these observables.

\begin{description}
\item[{\bf Thrust} \bm{T}:]  \hspace*{0pt\hfill} \\
     The thrust value is given by the expression~\cite{bib-thrust-def}
     \begin{displaymath}
            T = \max_{\vec{n}}\left( \frac{\sum_i | \vec{p}_i \cdot \vec{n} | }
                                          {\sum_i | \vec{p}_i | } \right)\ \ .
     \end{displaymath}
     The thrust axis $\vec{n}_T$ is the vector $\vec{n}$ which
     maximises the expression in parentheses.  It is used to divide an
     event into two hemispheres $H_1$ and $H_2$ by a plane through the
     origin and perpendicular to the thrust axis.
\item[{\bf Jet Broadening} \bm{B}:]  \hspace*{0pt\hfill} \\
     The jet broadening measures are calculated by~\cite{bib-NLLA-1}:
     \begin{displaymath}
          B_k = \left( \frac{\sum_{i\in H_k} | \vec{p}_i \times \vec{n}_T | }
                                         {2\sum_i | \vec{p}_i | } \right)
     \end{displaymath}
     for each of the two hemispheres, $H_k$, defined above.
     The total jet broadening is given $B_T = B_1 + B_2$. The wide
     jet broadening is defined by $B_W = \max(B_1, B_2)$.
\item[{\bf C-parameter}:]  \hspace*{0pt\hfill} \\
     The $C$-parameter is defined as~\cite{bib-C-parameter}
     \begin{displaymath}
       C = 3 (\lambda_1 \lambda_2  + \lambda_2 \lambda_3  + \lambda_3 \lambda_1 )
     \end{displaymath}
     where $\lambda_\gamma$, $\gamma=1, 2, 3$, are the eigenvalues of the momentum tensor
     \begin{displaymath}
       \Theta^{\alpha\beta} = \frac{\sum_i \vec{p}_i^{\,\alpha} \vec{p}_i^{\,\beta} / |\vec{p}_i|}
       {\sum_j |\vec{p}_j|}
       \ \ \ .
     \end{displaymath}
\end{description}

Experimental data for the thrust below the Z pole are provided e.g. by
the experiments of the PETRA (12 to 47~GeV), PEP (29~GeV) and TRISTAN
(55 to 58~GeV) colliders. For the thrust, we considered measured
distributions from the PETRA experiments JADE and TASSO (total of
about 50000 events), from the PEP experiments MARKII and HRS (total of
about 28000 events) and from the AMY detector (about 1200 events) at
TRISTAN (see Ref.~\cite{bib-data<91GeV}, and \cite{bib-newJADE}). For
the other observables, we used experimental distributions at PETRA
energies from our re-analysis of JADE
data~\cite{bib-newJADE,bib-newJADE2} providing about 22000 events at
35~GeV and 6200 events at 44~GeV.  The distributions used at the Z
resonance stem from a measurement using several hundred thousands
events for each of the four LEP experiments OPAL, ALEPH, DELPHI and
L3, and of about 40000 events for the SLD experiment at
SLAC~\cite{bib-data=91GeV}. We also considered measurements performed
by the LEP collaborations above the Z pole~\cite{bib-data>91GeV}
providing a few thousend events per
experiment up to a center-of-mass energy of 183~GeV%
%\footnote{ We point out that some of the LEP-2 data were presented in
%  previous HEP conferences but are still preliminary. We also used
%  these data although their statistical weight
%  to our final results is negligible.}
.%

All distributions used in this study were corrected for the limited
resolution and acceptance of the respective detectors and of the event
selection criteria (for details see the referenced publications).

\section{ Power corrections to differential distributions }
\label{sec-powcor}

Non-perturbative effects to event shape observables arise due to the
contributions of very low energetic gluons which cannot be described
perturbatively because of the presence of 
 an unphysical Landau pole in the
perturbative expression of the running coupling \asmz.  Within the
dispersive approach \cite{bib-webber,bib-Milan-factor}
a non-perturbative parameter
\begin{displaymath}
  \azeromuI = 
   \frac{1}{\muI} 
   \int_0^{\muI} {\mathrm{d}}k\ \ \as(k)
\end{displaymath}
is introduced to parametrize the analytic but unknown behaviour of
$\as(\rs)$ below a certain infrared matching scale \muI.
The technique presented in the cited papers also removes
the divergent soft gluon contributions in the perturbative predictions for
the observables and produces as a consequence 1/\rs\ corrections.

The power corrections which we used in this analysis have been
calculated in Ref.~\cite{bib-Milan-factor,bib-new-broadening} up to
two loops for the differential distributions of the event shapes
considered here.  The effect of hadronisation on the distribution
obtained from experimental data is described by a shift of the
perturbative prediction away from the kinematical 2-jet region,

\begin{equation}
  \frac{\mathrm{d}\sigma^{\mathrm{exp}}}{\mathrm{d}{\mathcal{F}}}\left(\mathcal{F}\right)  = 
  {\frac{\mathrm{d}\sigma^{\mathrm{pert}}}{\mathrm{d}{\mathcal{F}}}} \left ( \mathcal{F}- \mathcal{P} D_{\mathcal{F}} \right) \;.
\end{equation}
where $\mathcal{F}$ = $1-T$, $C$, $B_T$ and $B_W$, respectively.  The
factor $\mathcal{P}$ containing the only non-perturbative parameter
\azero\ is the same for all observables,
\begin{equation}
  \mathcal{P}  =  \frac{4C_F}{\pi^2} \mathcal{M} \frac{\muI}{\rs}
  \left( \alpha_0 (\muI) - \alpha_s(\rs) -\beta_0\frac{\alpha_s^2}{2\pi}
    \left( \ln \frac{\rs}{\muI}+\frac{K}{\beta_0}+1 \right) \right)
\end{equation}
where $C_F = 4/3$. $\beta_0 = (11C_A-2N_f)/3$ stems from the QCD
renormalisation group equation for $N_C=C_A$ colours and $N_f$ active
quark flavours. $K = (67/18-\pi^2)C_A-5/9N_f$ originates from the
choice of the $\overline{\mathrm{MS}}$ renormalisation scheme. The
Milan factor $\mathcal{M}$ accounts for two-loop effects.  Its
numerical value of 1.795 (for three active quark flavours) is given by
the authors of~\cite{bib-Milan-factor} with a theoretical uncertainty
of about 20\% owing to missing higher order corrections.
$D_{\mathcal{F}}$ is a function depending on the observable,
\begin{equation}
\label{eq-shift}
  D_{\mathcal{F}} = \left \{
    \begin{array}{ll}
      \mathrm{2}                        & ; { \mathcal{F} = 1-T}  \\[2mm]
      \mathrm{3\pi}                     & ; { \mathcal{F} = C}    \\[2mm]
      \frac{\s1}{\s2}\ln \frac{\s 1}{\s \mathcal{F}}+ B_{\mathcal{F}} (\mathcal{F},\as(\mathcal{F}\rs))  & ; { \mathcal{F} = B_T,\, B_W\;.}
    \end{array}   \right.
\end{equation}

Thus a simple shift is expected for thrust and $C$-parameter, whereas
for the jet broadening variables an additional squeeze of the
distribution is predicted. Besides the logarithmic dependence on the
value $\mathcal{F}$ of the observable, the term
$B_{\mathcal{F}}$ exhibits an additional (weaker) dependence on the
value of the observable and on the running coupling at the scale $\mathcal{F}\rs$.
The more complex behaviour for the jet broadening recently calculated
in~\cite{bib-new-broadening} is related to the interdependence of
non-perturbative and perturbative effects which cannot be neglected
for these observables.  As a consequence, the term
$\mathcal{P}D_{\mathcal{F}}$ leads not only to a larger shift but also
to a stronger squeeze at decreasing centre-of-mass energies. The
necessity of an additional non-perturbative squeeze of the jet
broadening distributions was already pointed out
in~\cite{bib-powcor-mp98}.

\section{ Tests of power corrections }
\label{sec-tests}
A few experimental tests of power corrections to differential
distributions have been done up to
now~\cite{bib-new-broadening,bib-powcor-mp98,bib-powcor-thrust-dokshitzer,bib-DELPHI-powcor-mp97,bib-ALEPH-powcor}.
Some studies for the new observables which were performed using LEP
data only suffer from the low data statistics above the Z pole
and from the lack of data at energies below the Z pole, where QCD scaling
violations and energy dependent non-perturbative effects are larger.
The
present analysis of combined LEP and PETRA data allows a more stringent test
of the $1/\rs$ dependence of the power corrections.

\subsection{ Fits to the data }
\label{subsec-fits}

The standard analysis uses the entire data set described in
Section~\ref{sec-data}. For the perturbative predictions, we employed
combined second order~\cite{bib-ERT} plus
resummed~\cite{bib-C-resummation,bib-new-jetbroadening,bib-NLLA-2,bib-NLLA-3}
QCD calculations (\oaa\ + NLLA). Predictions of this type take into
account the dominating leading and next-to-leading logarithms to all
orders in \as\ in the perturbation series for the observable and
therefore exhibit a weaker dependence on the QCD renormalisation scale
$\mu$ = $\xmu\rs$.  Since the matching procedure of the exact matrix
elements and the NLLA calculations reveals some ambiguities, we apply
the four different matching schemes proposed in~\cite{bib-NLLA-2}, the
R-, ln(R), the modified R and the modified ln(R)-matching.

For each observable, we performed simultaneous \chisq-fits of both
perturbative predictions plus power corrections which are parametrised
by \asmz\ and \azeromuI, respectively.  \asmz\ was evaluated to the
appropiate energy scale $\mu$ = $\xmu\rs$ using the two-loop formula for the
running coupling~\cite{bib-PDG}.  The errors used in the fit are the
quadratic sum of statistical and experimental systematic uncertainties
of the event shape distributions
denoted in the references. The errors of the fit are derived from
the 95\% CL two-dimensional contour.  The fit ranges are defined
individually for each center-of-mass energy such that they, in
general, include the maximum of the differential cross section located
in the 2-jet-region of the distributions.  In doing so we ensure to
exploit the range of the distribution at each energy point as far as
possible in order to constrain the theory  and to
consider as much data statistics as possible from the distributions,
which is in particular important in regard of the
scarce LEP-2 data.

\begin{figure}[!htb]
\vspace*{-10mm}\hspace*{15mm}
\begin{sideways}
\epsfig{file=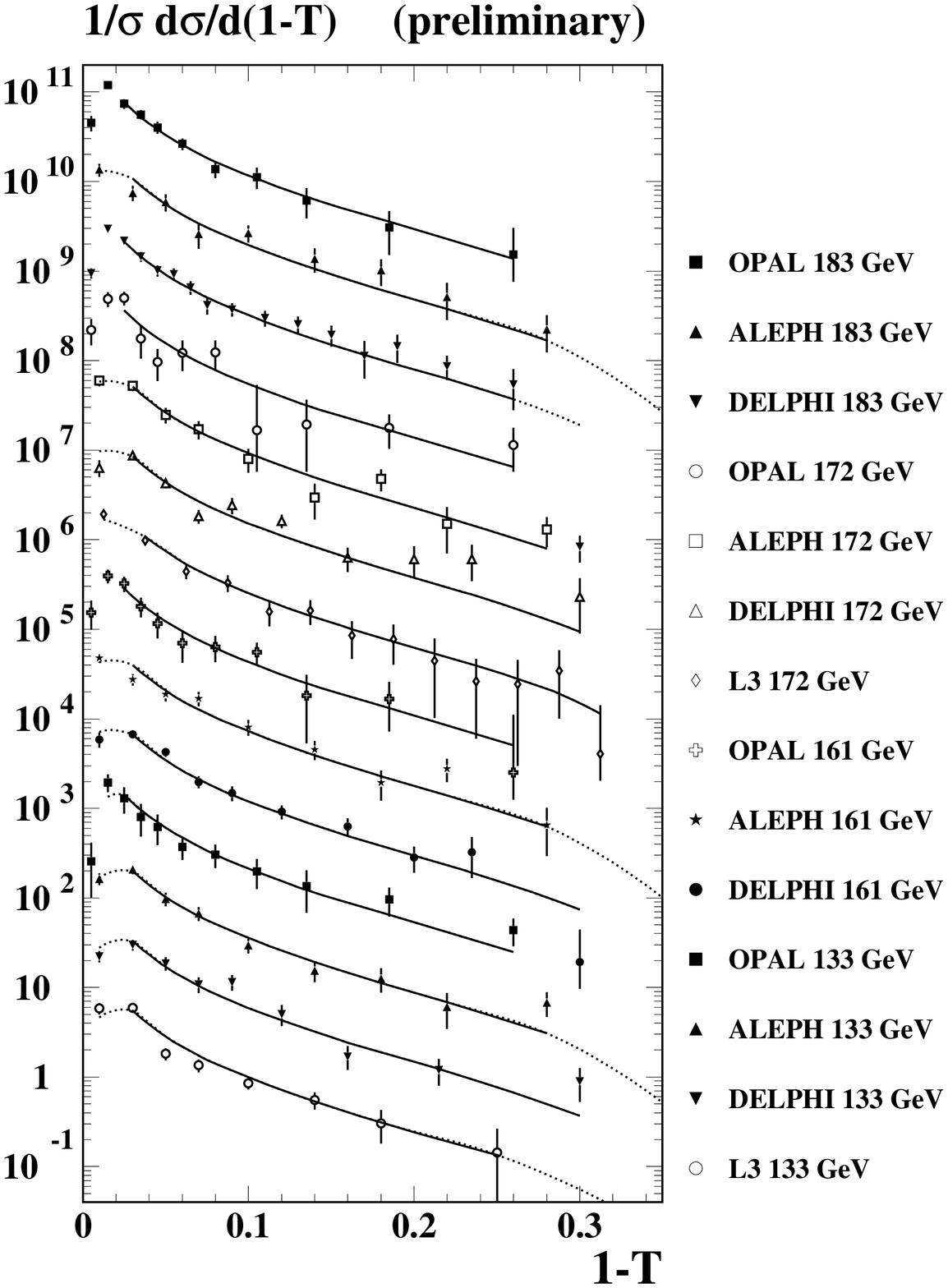,width=.6\textwidth,clip=}
\hspace*{-2mm}
\epsfig{file=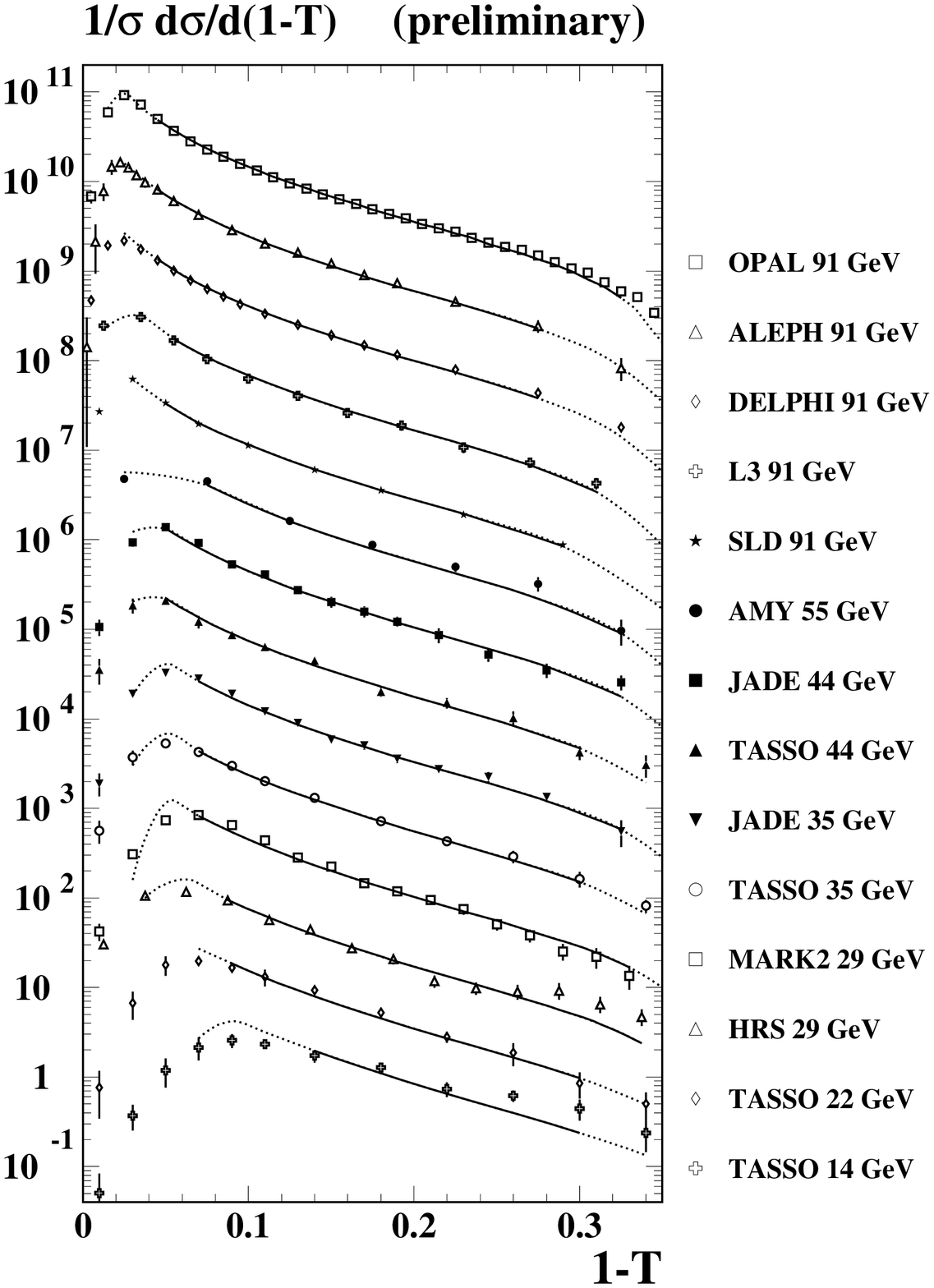,width=.6\textwidth,clip=}
\end{sideways}
\caption{ \label{fig-t_plot}
  Scaled distributions for $1-T$ measured by several experiments at
  $\protect\sqrt{s}= 14$ to $183$~GeV.  The error bars indicate the
  statistical and experimental errors of the data points.  The curves are the result
  of the simultaneous global (\as,\azero)-fit using
  resummed QCD predictions with the modified ln(R)-matching plus power
  corrections.  The fit ranges which are indicated by the solid lines
  are chosen individually for each center-of-mass energy.  }
\end{figure}

\begin{figure}[!htb]
\centerline{
\epsfig{file=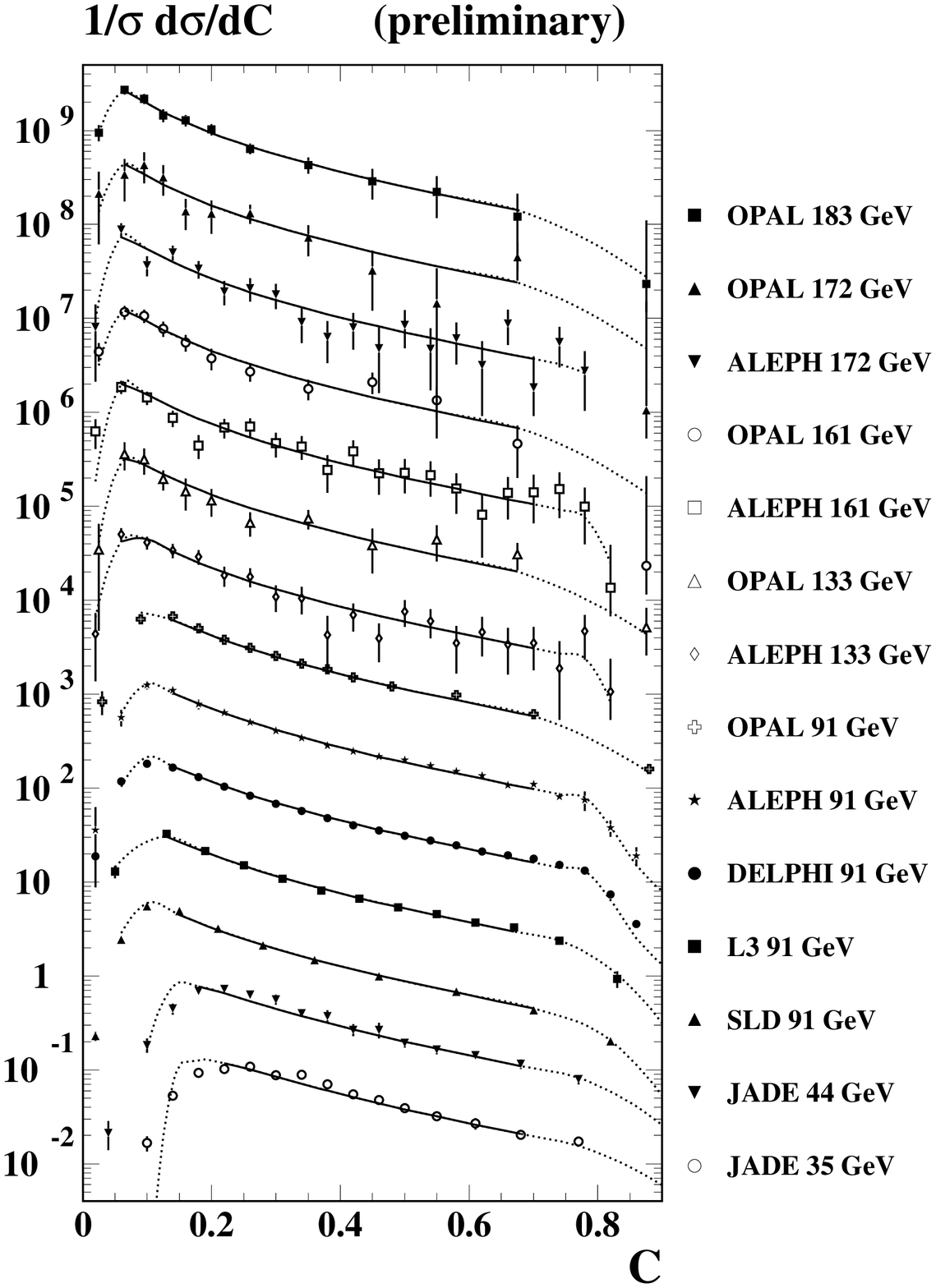,width=.7\textwidth,clip=}
}
\caption{ \label{fig-c_plot}
  Scaled distributions for the $C$-parameter measured by several
  experiments at $\protect\sqrt{s}= 35$ to $183$~GeV.  The error bars
  indicate the statistical and experimental errors of the data points.  The curves are
  the result of the simultaneous global (\as,\azero)-fit
  using resummed QCD predictions with the modified ln(R)-matching plus
  power corrections.  The fit ranges which are indicated by the solid
  lines are chosen individually for each center-of-mass energy.  }
\end{figure}

\begin{figure}[!htb]
\centerline{
\hspace*{8mm}
\epsfig{file=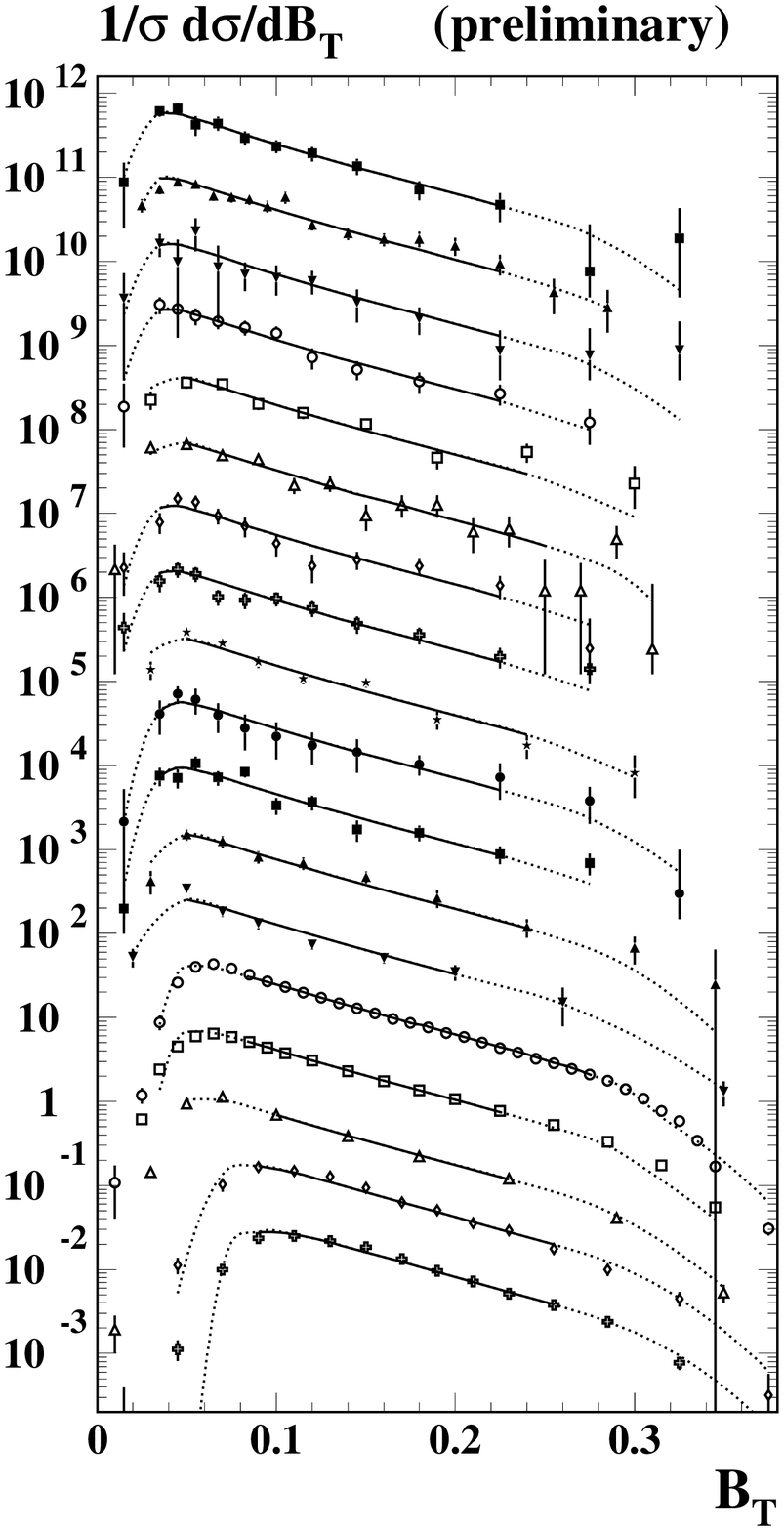,width=.7\textwidth,clip=}
\hspace*{-48.4mm}
\epsfig{file=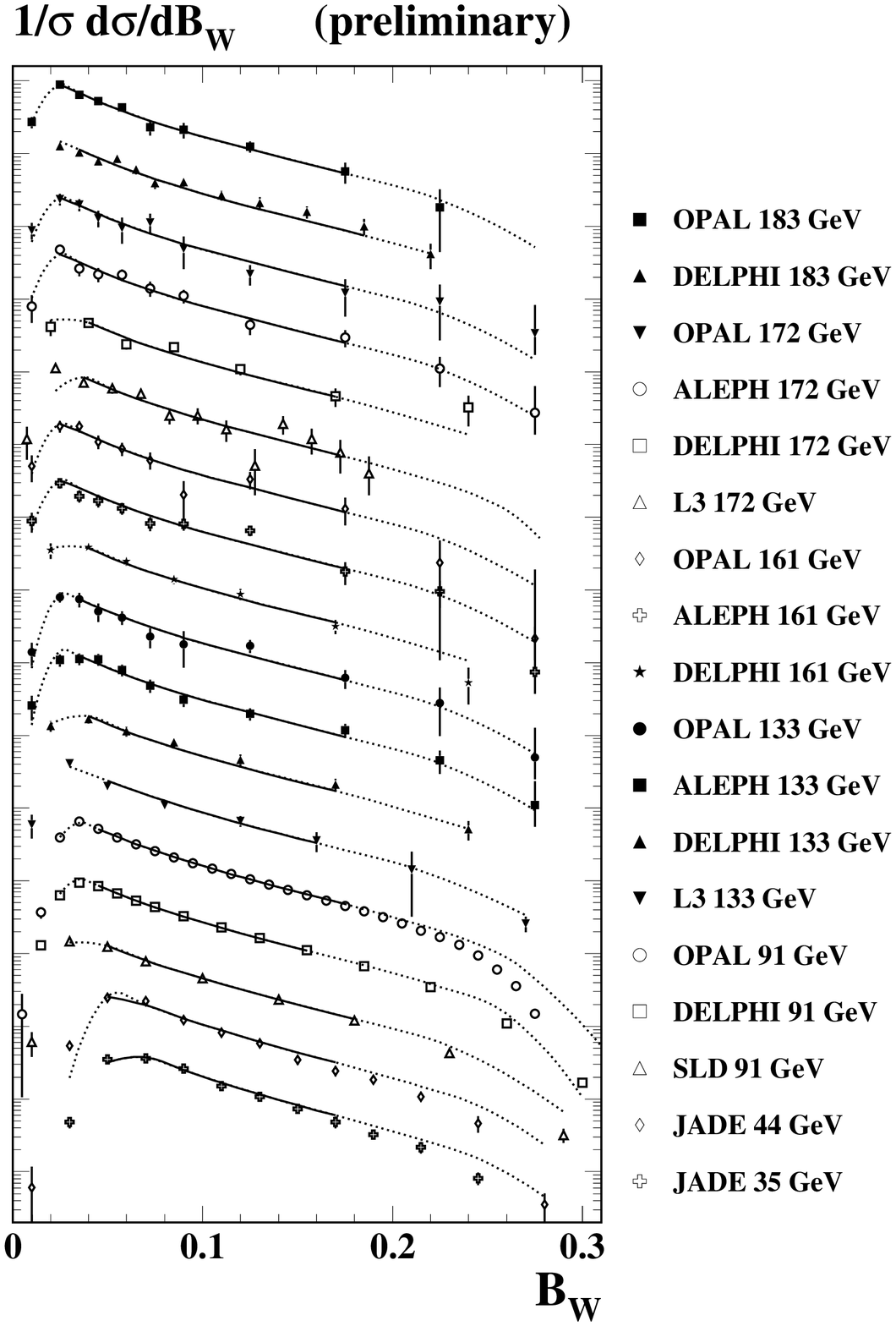,width=.7\textwidth,clip=}
}
\caption{ \label{fig-bt_bw_plot}
  Scaled distributions for \bt\ (left) and \bw\ (right) measured by
  several experiments at $\protect\sqrt{s}= 35$ to $183$~GeV.  The
  error bars indicate the statistical experimental errors of the data points.  The
  curves are the result of the simultaneous global
  (\as,\azero)-fit using resummed QCD predictions with the
  modified ln(R)-matching plus power corrections which include the
  revised power corrections to jet broadening
  observables\protect\cite{bib-new-broadening}.  The fit ranges which
  are indicated by the solid lines are chosen individually for each
  center-of-mass energy.  }
\end{figure}

Our standard result for an observable is defined by the unweighted
average of the results obtained from the different
\oaa+NLLA-combination schemes.  We fixed the renormalisation scale
\xmu\ = 1 and the infrared matching scale \muI\ = 2~GeV. The fit error
assigned is the average over the individual fit errors for the
matching schemes. The numeric results are listed in
Table~\ref{tab-as-powcor} for \as\ and in Table~\ref{tab-a0-powcor}
for $\alpha_0$. The fit curves for the modified ln(R)-matching and the
corresponding experimental data for the thrust $1-T$, the
$C$-parameter, the total and the wide jet broadening, \bt\ and \bw,
are shown in Figures~\ref{fig-t_plot}, \ref{fig-c_plot}
and~\ref{fig-bt_bw_plot}.  Generally we observe a good agreement of
the predictions with the data within the fit ranges, independent of
the matching scheme used in the fit.  The agreement is moderate for
some distributions at \rs\ $\le 35$~GeV.  The \chisqd\ of the fits
(see Table~\ref{tab-chisq}) are between 0.93 (for \bt, mod.
ln(R)-matching) and 1.34 (for \bw, R-matching). The fit quality
deteriorates for the jet broadenings if the upper limit of the fit
range is extended too far into the 3-jet region.  The excess of the
theory over the data observed here has also been noticed in other
studies~\cite{bib-OPALresummed,bib-DELPHIalphas,bib-SLDalphas} where
event shape distributions were corrected for hadronisation effects
using Monte Carlo models. Thus these discrepancies might be
due to the perturbative predictions.

The results for \asmz\ obtained from the fits proved to be
systematically lower than the respective results at \rs\ = \mz\ 
\cite{bib-newJADE2,bib-OPALresummed,bib-DELPHIalphas,bib-SLDalphas,bib-ALEPHalphas}
which use the same perturbative predictions but apply Monte Carlo
corrections instead of power corrections. They are consistent with
each other within the statistical errors in the case of thrust,
$C$-parameter and the total jet broadening \bt, whereas the results
obtained for \bw\ is about 20\% below the world average value for
\asmz~\cite{bib-world-alphas-sb}. This deviation is to a minor extend
also present in the studies cited.  Nevertheless, using the improved
power corrections the values for \asmz\ for \bt\ and \bw\ are now much
more compatible with the values obtained from the other observables
than is the case in our previous study~\cite{bib-powcor-mp98}.

The numeric results for \azerotwo\ vary between 0.437
for the $C$-parameter and 0.685 for \bw.  The high value for \bw\ is
supposed to be partially related to the systematically lower values of
\as, since the two fit parameters are anticorrelated (see
Table~\ref{tab-chisq} for the correlations coefficients).

From the experimental point of view, there is a lucid explanation for
the deviations of the \as\ results based on the power corrections from
those based on Monte Carlo corrections~\cite{bib-powcor-mp98}. We
realised in our studies that the latter induce a stronger squeeze to
{\em all} distributions than the power corrections which for instance
predict merely a shift for the thrust and the $C$-paramter without any
presence of a squeeze.  Although the situation was a little remedied
for the jet broadenings $B$ due to the factor ln $1/B$ appearing in
Equation~(\ref{eq-shift}), the extend of the squeezing remains below
the expectation of hadronisation models.  As a consequence, the
two-parameter fit favours systematically smaller values for \as\ in
order to make the predicted shape more peaked in the 2 jet region of
the distribution (thus causing a squeeze) and hence chooses high values
for \azero\ in order to compensate the resulting shift of the
distribution towards the 2-jet-region, as is in particular the case
for the jet broadenings.

\subsection{ Systematic uncertainties }
\label{subsec-systematics}
In order to scrutinize the universality of the non-perturbative
parameter \azero\, we also studied the impact of systematic
uncertainties to our final results. Theoretical uncertainties from
perturbation theory were assessed by considering the different
matching schemes and by varying \xmu\ from 0.5 to 2.0.  For the
further systematic checks, we took the modified ln(R)-matching scheme
as a reference since it turned out that most of the relative changes
of \as\ and \azero\ are not significantly affected by the choice of
the matching scheme.  Uncertainties due to the parametrisation of the
power corrections are associated with the arbitrariness of the choice
of the value of \muI\ which has been varied by $\pm$ 1~GeV.
Furthermore the Milan factor $\mathcal{M}$ was varied within the
quoted uncertainty of 20\%.  No error contribution from the infrared
matching scale is assigned to \azeromuI.

The signed values in Tables~\ref{tab-as-powcor}
and~\ref{tab-a0-powcor} indicate the direction in which \asmz\ and
\azero\ changed with respect to the standard analysis.  It turns
out that the non-perturbative uncertainties for \asmz\ are negligible
for each observable. This is in accordance with the expectation since
\as\ is mainly constrained by the perturbative prediction and not by
the non-perturbative shift %
$\propto \muI\mathcal{M}(\azero-\as+\oaa)$. %
On the other hand, the strong dependence of \azero\ on $\mathcal{M}$
is due to the obvious  anticorrelation.  In regard to the
remarks made in Section~\ref{subsec-fits} it is noticable that the
value for \azero\ from \bw\ is affected with the largest
non-perturbative error.

We also examined the dependence of the results from the input data
taken for the fits which is for brevity denoted as ``experimental
uncertainty''. Since the statistical weight of the event shape
distributions at \rs\ = 91~GeV is dominant in the fits, the analysis
was repeated using the data from each LEP/SLC experiment seperately as
representative for \rs~= 91~GeV. Each deviation from the standard
result was added in quadrature. In addition, we performed a
measurement omiting the 91~GeV data.  The larger of the two latter
deviations were incorporated in the experimental error.  Furthermore, an
analysis was performed using seperately either event shape data below
the Z pole or data at \rs~$\ge M_{Z^0}$. This checks also for higher
order non-perturbative contributions $\propto (1/\rs)^2$ to the power
corrections.

A further source of experimental uncertainty comes from the choice of
the fit range. We quantified it by varying the lower and the upper
edge of the fit range of all distributions likewise in both directions
by an amount approximately corresponding to the respective bin of the
JADE distribution at 35~GeV. The fit range of a distribution was
expanded into the 2-jet region as far as the \chisq\ of
the extreme bin did not
contribute substantially to the total \chisq\ of the distribution.
 For each edge, we took the larger deviation from the standard
result as error.

In general, the experimental errors are more moderate than the
theoretical uncertainties. We realise major variations of \asmz\ from
\bt\ and \bw\ when considering only PETRA data in the fits. The larger
dependence of the results on the lower edge of the fit range of the thrust
distribution may be partially ascribed to the contributions of the lower energy data
down to 14~GeV at which mass effects become important.

All errors are treated as asymmetric uncertainties on \asmz\ and
\azeromuI, respectively. The total errors are quoted as the
quadratic sum of the individual uncertainties, thus disregarding
possible correlations between them.  They are also given in
Table~\ref{tab-as-powcor} and \ref{tab-a0-powcor} for \as\ and
\azero, respectively.

\subsection{ Combination of individual results }
\label{sec-comb}
The individual results for the four event shape observables were
combined to a single value for \asmz\ and \azerotwo,
respectively, following the procedure described in~\cite{bib-newJADE}.
This procedure accounts for correlations of the systematic errors. A
weighted average of the individual results was calculated with the
square of the reciprocal total errors used as the
weight. For each of the systematic checks, the mean values for \asmz\ 
and \azerotwo\ from all considered
observables were determined. Any deviation from the weighted average
of the main result was taken as a systematic uncertainty.

\begin{figure}[!htb]
\centerline{
 \hspace*{-7mm}
 \epsfig{file=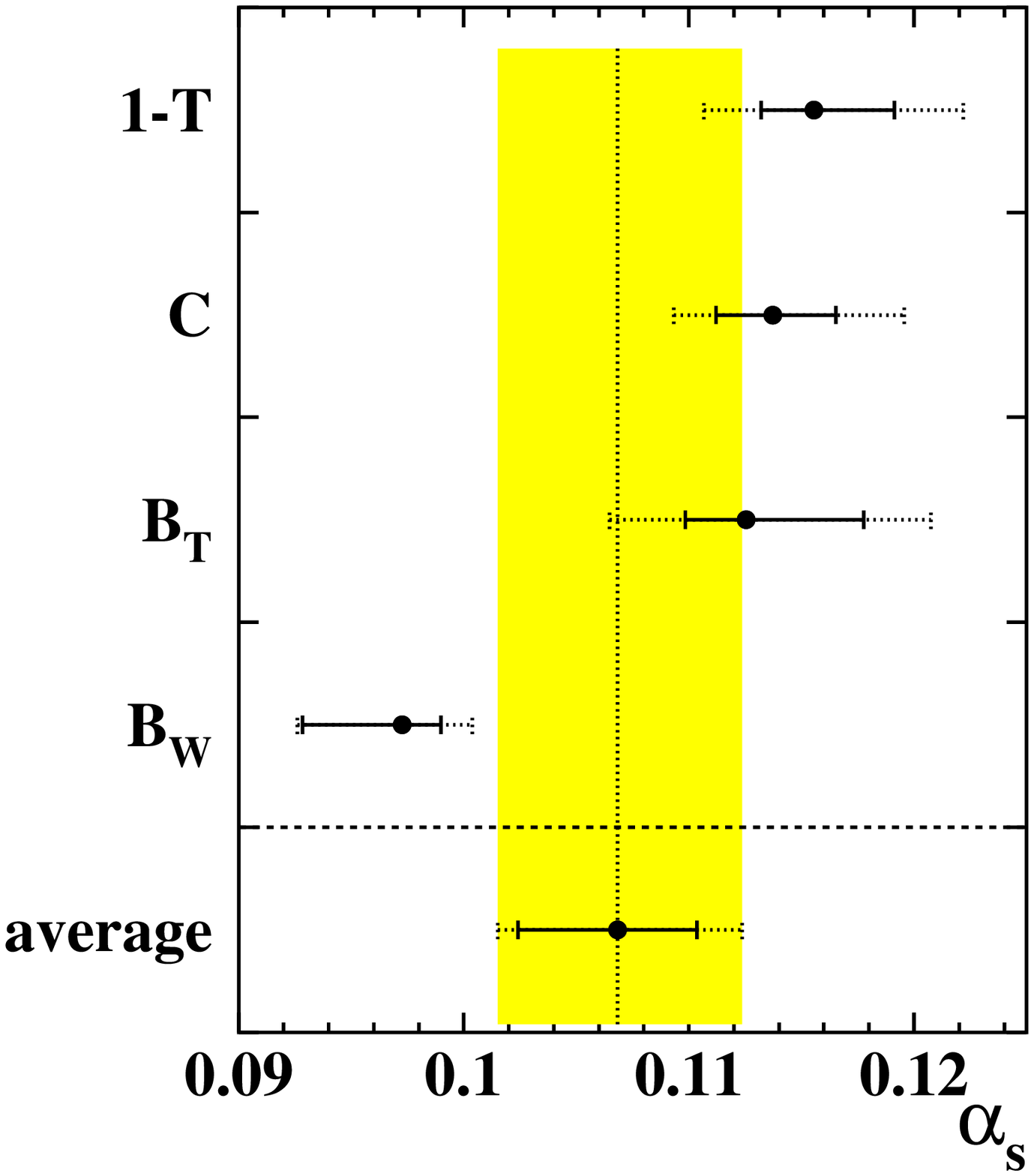,width=.6\textwidth,clip=}
 \hspace*{-31.5mm}
 \epsfig{file=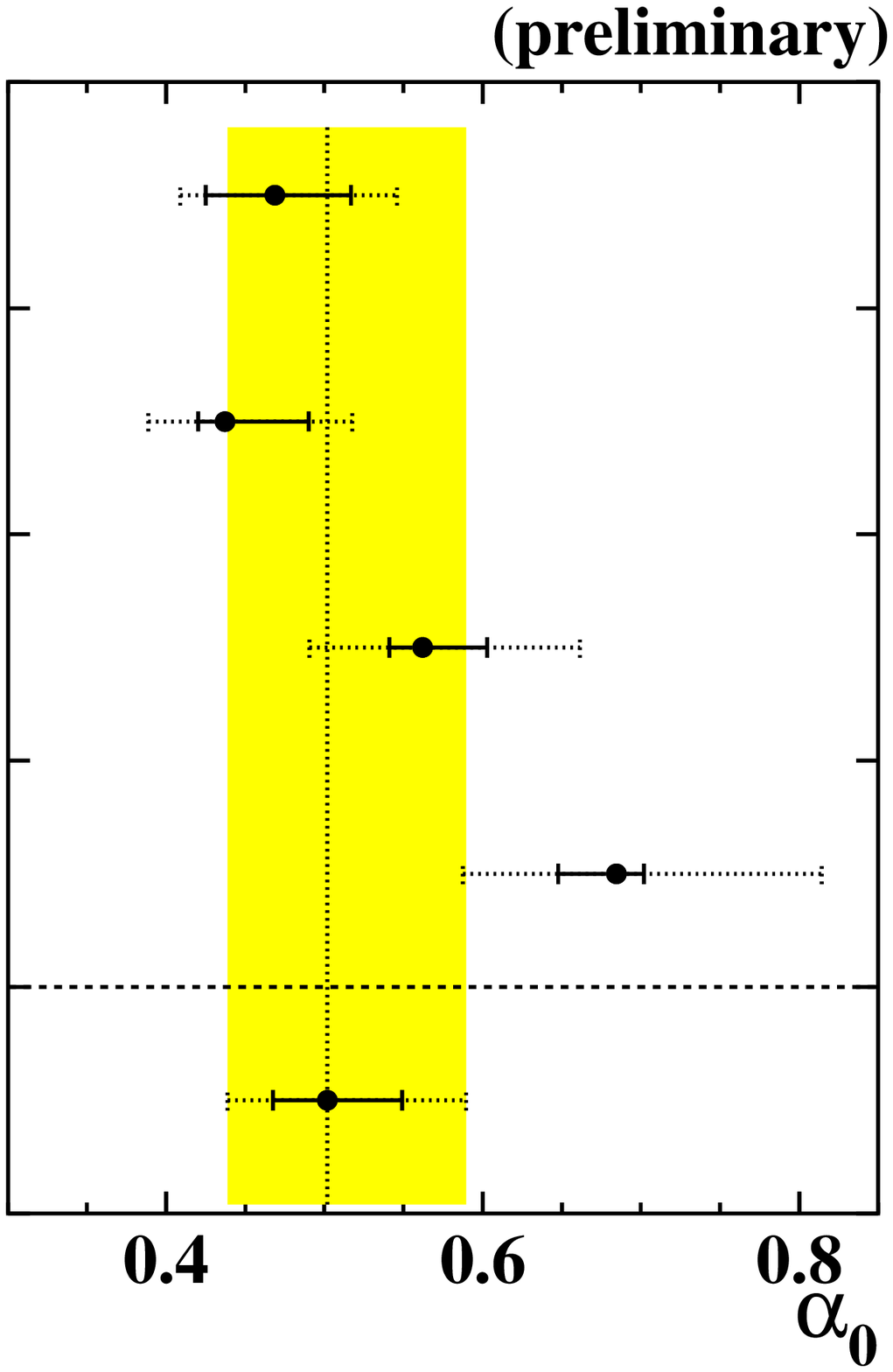,width=.6\textwidth,clip=}
}
\caption{ \label{fig-a0_results}
  Results for \asmz\ and \azerotwo\ from fits of power corrections 
  to the event shape distributions of thrust $T$, $C$-parameter, total
  and wide jet broadening, \bt\ and \bw, respectively.
  The experimental and statistical uncertainties
  are represented by the solid error bars. The dashed error bars show the
  total error including theoretical uncertainties from the resummed QCD
  and the power corrections. The shaded band shows the one standard
  deviation around the weighted average.
}
\end{figure}

With this procedure, we obtained as final result for \azerotwo\ 
\begin{eqnarray*}
\azerotwo
       =  0.502 \pm 0.013 \mathrm{(stat.)}
                         \ ^{+0.046} 
                           _{-0.032} \mathrm{(exp. \; syst.)}
                         \ ^{+0.074} 
                           _{-0.053} \mathrm{(theo. \; syst.)}
\end{eqnarray*}
where the error is dominated by non-perturbative uncertainties.
Figure~\ref{fig-a0_results} illustrates that the individual results
are compatible with the weighted mean within the total errors. We
consider this result as a confirmation of the predicted universality
of the non-perturbative parameter \azero.

Employing the same procedure for \asmz, we obtain
\begin{eqnarray*}
 \asmz = 0.1068  \pm 0.0011 \mathrm{(stat.)}
                 \ ^{+0.0033}
                   _{-0.0043} \mathrm{(exp. \; syst.)}
                   ^{+0.0043}
                   _{-0.0029} \mathrm{(theo. \; syst.)} \;.
\end{eqnarray*}
This small value compared to the world average~\cite{bib-world-alphas-sb}
is due to the \as\ from \bw\ which is substantially below 
the \as\ from $T$, $C$-parameter \bt\ (see Figure~\ref{fig-a0_results}).

If \bw\ is omited, the final value is
$\asmz =  0.1141$ 
$\pm 0.0012 \mathrm{(stat.)}$ 
                         $ ^{+0.0034}
                           _{-0.0024} \mathrm{(exp. \; syst.)}$
                         $ ^{+0.0055}  
                           _{-0.0041} \mathrm{(theo. \; syst.)}$
which is in good agreement with the world average value~\cite{bib-world-alphas-sb}.
In this case, the result for \azero\ does insignificantly change to
$\azerotwo = 0.478$
           $\pm 0.013 \mathrm{(stat.)}$
                        $ ^{+0.047} 
                           _{-0.020} \mathrm{(exp. \; syst.)}$
                        $ ^{+0.067} 
                           _{-0.048} \mathrm{(theo. \; syst.)}$.

\section{ Summary and conclusions }
\label{sec-conclusions}
The analytic treatment of non-perturbative effects to hadronic event
shapes based on power corrections were examined. We tested
predictions~\cite{bib-Milan-factor,bib-new-broadening} for the
differential distributions of thrust, $C$-parameter, total and wide
jet broadening, \bt\ and \bw, respectively. For this test, a large
amount of event shape data collected by several experiments over a huge
range of \epem annihilation energies from \rs~= 14 to 183~GeV was
considered. For the $C$-parameter, \bt\ and \bw, differential 
distributions below the Z pole are provided by our re-analysis
of \epem data of the JADE experiment~\cite{bib-newJADE,bib-newJADE2}
at \rs~= 35 and 44~GeV.

A simultaneous two-parameter fit of the combined power corrections
plus perturbative QCD-predictions (\oaa+NLLA) was performed for both
the strong coupling \asmz\ and the non-perturbative parameter
\azeromuI.  The good quality of the fits with \chisqd\ around 1.0 for
each observable supports the predicted $1/\rs$-evolution of the power
corrections.  The results for \asmz\ and \azerotwo\ are more
consistent with each other due to the improved predictions of power
corrections to the jet broadening variables~\cite{bib-new-broadening}.
Nevertheless we still observe a large deviation of the results
obtained from \bw\ from those extracted from the other observables.
We suspect it to be partially due to the resummed predictions for this
observable.  The individual results for \as\ from all observables
(Table~\ref{tab-as-powcor}) are systematically smaller than the
respective results
in~\cite{bib-newJADE2,bib-OPALresummed,bib-DELPHIalphas,bib-SLDalphas,bib-ALEPHalphas},
the latter employing hadronisation models for the correction of
non-perturbative effects. This observation is presumably related to
the different extends of the non-perturbative squeeze of the
distributions predicted by the both ansatzes.
We obtain as combined result
\[ \asmz = 0.107 \ ^{+0.006}_{-0.005}\; . \]
It should be noted that \as\ from the wide jet broadening \bw\ 
is rather small compared with the \as\ from the other observables.
 The combined value for
\asmz\ derived when omiting the results for \bw\ is $\asmz = 0.114 \ 
^{+0.007}_{-0.005}(\mathrm{tot.})$, in agreement with direct
measurements at the Z
resonance~\cite{bib-OPALresummed,bib-DELPHIalphas,bib-SLDalphas,bib-ALEPHalphas}
and with the world average value of $\as^{w.a.}(M_{\mathrm{Z^0}}) =
0.119 \pm 0.004$.

The combination of the individual results for \azero\ from all
observables considered yields as final result for the coupling moment
\[ \alpha_0(2~\mathrm{GeV}) = 0.50 \ ^{+0.09}_{-0.06}\]
where the error is dominated by theoretical uncertainties defined in
Section~\ref{subsec-systematics}. Since this value is representative
for the individual results within the errors defined, we consider this
as a confirmation of the universality of \azero\ as predicted within
the dispersive approach~\cite{bib-Milan-factor} of power corrections.
Nevertheless, the results yielded from \bt\ and \bw\ indicate that
higher orders may contribute significantly to the non-perturbative
corrections for the jet broadenings.

We point out that the average value for \azerotwo\ derived from the
differential event shape distributions is in good agreement with the
result of $\azerotwo = 0.47 \ ^{+0.06}_{-0.04}$ provided by our
previous study~\cite{bib-newJADE2} of power corrections to the mean
values of event shapes.

%%%%%%%%%%%% Start standard additions %%%%%%%%%%%%%%%%%%%
\medskip
\bigskip\bigskip\bigskip
\appendix
\par
\section*{Acknowledgements}
\par
We are grateful to G.~Salam for helpful discussions.

%%%%%%%%%%%%%%%%%%%%%%%%%%%%%%%%%%%%%%%%%%%%%%%%%%%%%%%%%%%%%%%%%%%%%%%%%%%
%  The tables:
\clearpage
\section*{ Tables }

%
% Table of alpha_s results
%
\begin{table}[!htb]
\begin{center}

\begin{tabular}{|r||c|c|c|c|}%|c|}
\hline
                   &$\mathbf{1-T}$        &$\mathbf{C}$        &$\mathbf{B_T}$        &$\mathbf{B_W}$ \\%& average\\
\hline\hline
$\mathbf{\asmz}$     &\bf  0.1156     &\bf  0.1137     &\bf  0.1125     &\bf 0.0973    \\% &\bf  0.1069     \\ %
\hline%.
\bf Exp. stat.     &\bf $\pm$ 0.0011     &\bf $\pm$ 0.0011     &\bf $\pm$ 0.0014     &\bf $\pm$ 0.0009    \\% &\bf $\pm$ 0.0011     \\ %
%\hline%.
%\bf $\chi^2$/dof   &\bf   306.1/ 277 %
%                             &\bf   169.1/ 170 %
%                                       &\bf   164.9/ 171 %
%                                                 &\bf   152.9/ 132\\%& \\ %
\hline\hline%.
\raisebox{.5ex}[-.5ex]{\bf pQCD  }  & $\stackrel{\s\mathbf{+0.0055}}{\s\mathbf{-0.0043}}$      %
                             & $\stackrel{\s\mathbf{+ 0.0051}}{\s\mathbf{-0.0036}}$      %
                                       & $\stackrel{\s\mathbf{+ 0.0063}}{\s\mathbf{-0.0054}}$      %
                                                 & $\stackrel{\s\mathbf{+ 0.0026}}{\s\mathbf{-0.0015}}$   \\%   %
%                                                 & $\stackrel{\s\mathbf{+ 0.0043}}{\s\mathbf{-0.0030}}$     \\ %
\hline%.
\raisebox{.5ex}[-.5ex]{\bf Pow. corr. }    & $\stackrel{\s\mathbf{+ 0.0007}}{\s\mathbf{-0.0006}}$      %
                             & $\stackrel{\s\mathbf{+ 0.0000}}{\s\mathbf{-0.0001}}$      %
                                       & $\stackrel{\s\mathbf{+ 0.0002}}{\s\mathbf{-0.0003}}$      %
                                                 & $\stackrel{\s\mathbf{+ 0.0000}}{\s\mathbf{-0.0000}}$    \\%  %
%                                                 & $\stackrel{\s\mathbf{+ 0.0002}}{\s\mathbf{-0.0002}}$     \\ %
\hline%.
\raisebox{.5ex}[-.5ex]{\bf Exp. syst. }    & $\stackrel{\s\mathbf{+ 0.0034}}{\s\mathbf{-0.0021}}$      %
                             & $\stackrel{\s\mathbf{+ 0.0026}}{\s\mathbf{-0.0022}}$      %
                                       & $\stackrel{\s\mathbf{+ 0.0050}}{\s\mathbf{-0.0023}}$      %
                                                 & $\stackrel{\s\mathbf{+ 0.0015}}{\s\mathbf{-0.0043}}$     \\% %
%                                                 & $^{\s+ 0.0033}_{\s-0.0043}$     \\ %
\hline\hline%.
\raisebox{.5ex}[-.5ex]{\bf Total  }        & $\stackrel{\s\mathbf{+ 0.0066}}{\s\mathbf{-0.0049}}$      %
                             & $\stackrel{\s\mathbf{+ 0.0058}}{\s\mathbf{-0.0044}}$      %
                                       & $\stackrel{\s\mathbf{+ 0.0082}}{\s\mathbf{-0.0061}}$      %
                                                 & $\stackrel{\s\mathbf{+ 0.0031}}{\s\mathbf{-0.0046}}$     \\% %
%                                                 & $^{\s+ 0.0054}_{\s-0.0053}$     \\ %
\hline\hline%.
ln(R)              &$ -0.0002   $&$ -0.0001   $&$ -0.0015   $&$ \lt0.0001  $  \\% %
%                                                 & $^{\s+ \lt0.0001}_{\s-0.0004}$     \\ %
\hline%.
ln(R) mod.         &$ +0.0008   $&$+0.0011    $&$+0.0029    $&$+0.0013 $   \\% %
%                                                $& $^{\s+ 0.0014}_{\s \lt0.0001}$     \\ %
\hline%.
R                  &$ -0.0010   $&$ -0.0005   $&$ -0.0032   $&$ -0.0010 $   \\% %
%                                                 & $^{\s+ \lt0.0001}_{\s-0.0012}$     \\ %
\hline%.
R mod.             &$ +0.0004   $&$ -0.0005   $&$ +0.0018   $&$ -0.0003  $  \\% %
%                                                 & $^{\s+ 0.0009}_{\s-0.0004}$     \\ %
\hline%.
$x_{\mu}=0.5$      &$ -0.0041   $&$ -0.0035   $&$ -0.0041   $&$ -0.0010  $  \\% %
%                                                 & $^{\s+ \lt0.0001}_{\s-0.0027}$     \\ %
\hline%.
$x_{\mu}=2.0$      &$ +0.0055   $&$ +0.0050   $&$ +0.0053   $&$ +0.0022 $  \\%  %
%                                                 & $^{\s+ 0.0039}_{\s \lt0.0001}$     \\ %
\hline\hline%.
${\cal M}- 20\% $  &$ +0.0003   $&$  \lt0.0001   $&$ +0.0002  $&$  \lt0.0001 $   \\% %
%                                                 & $^{\s+ 0.0001}_{\s \lt0.0001}$     \\ %
\hline%.
${\cal M}+ 20\% $  &$ -0.0003   $&$  \lt0.0001   $&$ -0.0001  $&$  \lt0.0001 $   \\% %
%                                                 & $^{\s+ \lt0.0001}_{\s-0.0002}$     \\ %
\hline\hline%.
91 GeV = OPAL      &$ -0.0006   $&$ +0.0004   $&$ -0.0011   $&$ -0.0023  $  \\% %
%                                                 & $^{\s+ 0.0004}_{\s-0.0016}$     \\ %
\hline%.
91 GeV = ALEPH     &$ -0.0001   $&$ -0.0004   $&    ---     &    ---      \\% %
%                                                 & $^{\s+ \lt0.0001}_{\s-0.0003}$     \\ %
\hline%.
91 GeV = DELPHI    &$ -0.0001   $&$ +0.0005   $&$ +0.0009   $&$ +0.0011  $  \\% %
%                                                 & $^{\s+ 0.0009}_{\s-0.0001}$     \\ %
\hline%.
91 GeV = L3        &$ -0.0010   $&$ -0.0011   $&    ---      &    ---      \\% %
%                                                 & $^{\s+ \lt0.0001}_{\s-0.0010}$     \\ %
\hline%.
91 GeV = SLD       &$ -0.0014   $&$ -0.0017   $&$ +0.0004   $&$ -0.0008  $  \\% %
%                                                 & $^{\s+ 0.0004}_{\s-0.0012}$     \\ %
\hline%.
$\rs \ne 91$ GeV   &$ -0.0003   $&$ -0.0004   $&$ +0.0020   $&$ -0.0012  $  \\% %
%                                                 & $^{\s+ 0.0020}_{\s-0.0008}$     \\ %
\hline%.
$\rs \ge 91$ GeV   &$ +0.0019   $&$  \lt0.0001  $&$ +0.0002   $&$ +0.0001 $   \\% %
%                                                 & $^{\s+ 0.0005}_{\s \lt0.0001}$     \\ %
\hline%.
$\rs < 91$ GeV     &$ +0.0007   $&$ +0.0024   $&$ +0.0046   $&$ -0.0035  $  \\% %
%                                                 & $^{\s+ 0.0023}_{\s-0.0035}$     \\ %
\hline%.
\raisebox{.5ex}[-.5ex]{fit range}           & $\stackrel{\s+ 0.0027}{\s-0.0010}$      %
                             & $\stackrel{\s+ 0.0004}{\s-0.0009}$      %
                                       & $\stackrel{\s+ 0.0006}{\s-0.0021}$      %
                                                 & $\stackrel{\s+ 0.0010}{\s-0.0005}$  \\%    %
%                                                 & $^{\s+ 0.0012}_{\s-0.0009}$     \\ %
\hline%.
\end{tabular}
\end{center}
\caption{\label{tab-as-powcor} 
  Values of \asmz\ derived using combined power corrections plus
  resummed QCD predictions to the event shape
  distributions of trust $T$, $C$-parameter, total
  and wide jet broadening \bt\ and \bw. The standard
  result for each variable is defined as the unweighted average of the
  results from the different matching schemes
  (ln(R), ln(R) mod., R, R mod.) using
  $\xmu=1$ and $\muI=2$~GeV and considering the entire data set.
  In addition, the statistical and  systematic uncertainties are given. 
  Signed values indicate the direction in which  \asmz\ changed 
  with respect to the standard analysis.
  Theoretical uncertainties are given by the matching scheme ambiguity,
  the choice of $\xmu$ and  $\muI$ and the uncertainty 
  of the Milan factor ${\cal M}$. 
  Experimental uncertainties are taken into account by varying the
  input data of the fit (see text).
  All errors  are treated as asymmetric uncertainties on \asmz. 
}
\end{table}

\newpage
%
% Table of alpha_0 results
%
\begin{table}[!htb]
\begin{center}
\begin{tabular}{|r||c|c|c|c|}%|c|}
\hline
                   &$\mathbf{1-T}$    &$\mathbf{C}$  &$\mathbf{B_T}$ &$\mathbf{B_W}$\\% & average\\
\hline\hline
$\mathbf{\azero}$&\bf  0.469      &\bf  0.437      &\bf  0.562      &\bf  0.685   \\%   &\bf  0.5008     \\ %
\hline%.
\bf Exp. stat.     &\bf $\pm$ 0.011      &\bf $\pm$ 0.010      &\bf $\pm$ 0.017      &\bf $\pm$ 0.014   \\%   &\bf $\pm$ 0.0128     \\ %
\hline%.
%$\chi^2$/dof   &\bf   306.1/ 277 %
%                             &\bf   169.1/ 170 %
%                                       &\bf   164.9/ 171 %
%                                                 &\bf   152.9/ 132\\%& \\ %
%\hline\hline%.
 \raisebox{.5ex}[-.5ex]{\bf pQCD }          & $\stackrel{\s\mathbf{+ 0.009 }}{\s\mathbf{-0.009 }}$      %
                             & $\stackrel{\s+ \mathbf{0.026 }}{\s\mathbf{-0.027 }}$      %
                                       & $\stackrel{\s\mathbf{+ 0.041 }}{\s\mathbf{-0.041 }}$      %
                                                 & $\stackrel{\s\mathbf{+ 0.029 }}{\s\mathbf{-0.033 }}$  %
                                                          \\%
\hline%.
 \raisebox{.5ex}[-.5ex]{\bf Pow. corr. }     & $\stackrel{\s\mathbf{+ 0.060 }}{\s\mathbf{-0.040 }}$      %
                             & $\stackrel{\s\mathbf{+ 0.055 }}{\s\mathbf{-0.037 }}$      %
                                       & $\stackrel{\s\mathbf{+ 0.081 }}{\s\mathbf{-0.054 }}$      %
                                                 & $\stackrel{\s\mathbf{+ 0.125 }}{\s\mathbf{-0.084 }}$  %
                                                         \\%
\hline%.
 \raisebox{.5ex}[-.5ex]{\bf Exp. syst.}    & $\stackrel{\s\mathbf{+ 0.047 }}{\s\mathbf{-0.042 }}$      %
                             & $\stackrel{\s\mathbf{+ 0.052 }}{\s\mathbf{-0.013 }}$      %
                                       & $\stackrel{\s\mathbf{+ 0.037 }}{\s\mathbf{-0.012 }}$      %
                                                 & $\stackrel{\s\mathbf{+ 0.011 }}{\s\mathbf{-0.034 }}$ %
                                                         \\%
\hline\hline%.
 \raisebox{.5ex}[-.5ex]{\bf Total }  & $\stackrel{\s\mathbf{+ 0.077 }}{\s\mathbf{-0.060 }}$      %
                             & $\stackrel{\s\mathbf{+ 0.081 }}{\s\mathbf{-0.049 }}$      %
                                       & $\stackrel{\s\mathbf{+ 0.099 }}{\s\mathbf{-0.071 }}$      %
                                                 & $\stackrel{\s\mathbf{+ 0.130 }}{\s\mathbf{-0.097 }}$   %
                                                         \\%
\hline\hline%.
ln(R)              &$+0.003    $&$ +0.019   $&$ +0.023   $&$+0.012   $  \\% &+0.0134     \\%
\hline%.
ln(R) mod.         &$ -0.005   $&$ -0.015   $&$ -0.025   $&$ -0.019  $   \\% & -0.0139     \\%
\hline%.
R                  &$ +0.007   $&$ +0.016   $&$ +0.033   $&$+0.021   $  \\% &+0.0167     \\%
\hline%.
R mod.             &$ -0.005      $&$ -0.020      $&$ -0.032   $&$ -0.014  $   \\% & -0.0163     \\%
\hline%.
$x_{\mu}=0.5$      &$ -0.006      $&$ -0.012      $&$ -0.008   $&$ -0.023  $  \\%  & -0.0102     \\%
\hline%.
$x_{\mu}=2.0$      &$+0.005      $&$+0.008      $&$+0.007      $&$+0.017  $  \\%  &+0.0074     \\%
\hline\hline%.
${\cal M}- 20\% $  &$+0.060      $&$+0.055      $&$+0.081      $&$  +0.125  $   \\% &+0.0696     \\%
\hline%.
${\cal M}+ 20\% $  &$ -0.040      $&$ -0.037      $&$ -0.054   $&$ -0.084  $  \\%  & -0.0466     \\%
\hline\hline%.
91 GeV = OPAL      &$+0.018      $&$+0.027      $&$+0.022      $&$ +0.010  $   \\% &+0.0209     \\%
\hline%.
91 GeV = ALEPH     &$+0.018      $&$+0.020      $&    ---       &    ---    \\%   &+0.0189     \\%
\hline%.
91 GeV = DELPHI    &$ <0.001      $&$ -0.013      $&$ -0.010    $&$ -0.017  $ \\%   & -0.0083     \\%
\hline%.
91 GeV = L3        &$+0.021      $&$+0.024      $&    ---       &    ---   \\%    &+0.0224     \\%
\hline%.
91 GeV = SLD       &$+0.020      $&$+0.025      $&$+0.006      $&$ -0.024  $  \\%  &+0.0138     \\%
\hline%.
$\rs \ne 91$ GeV   &$+0.020      $&$+0.021      $&$+0.008      $&$ -0.021  $ \\%   &+0.0132     \\%
\hline%.
$\rs \ge 91$ GeV   &$ -0.029      $&$ -0.003      $&$ -0.005   $&$+0.005  $  \\%  & -0.0115     \\%
\hline%.
$\rs < 91$ GeV     &$+0.020    $&$+0.019      $&$+0.004      $&$+0.001   $ \\%  &+0.0141     \\%
\hline%.
 \raisebox{.5ex}[-.5ex]{fit range }         & $\stackrel{\s+ 0.018 }{\s-0.031 }$      %
                             & $\stackrel{\s+ 0.008 }{\s-0.001 }$      %
                                       & $\stackrel{\s+ 0.030 }{\s-0.005 }$      %
                                                 & $\stackrel{\s+ 0.002 }{\s-0.017 }$     \\ %
\hline%.
\end{tabular}
\end{center}
\caption{\label{tab-a0-powcor} 
  Values of \azeromuI\ derived using combined power corrections plus
  resummed QCD predictions to the event shape
  distributions of trust $T$, $C$-parameter, total
  and wide jet broadening \bt\ and \bw. The standard
  result for each variable is defined as the average of the
  results from the different matching schemes
  (ln(R), ln(R) mod., R, R mod.) using $\xmu$~= 1 and $\muI$ =2~GeV
  and considering the entire data set.
  In addition, the statistical and systematic uncertainties are given. 
  Signed values indicate the direction in which  \asmz\ changed 
  with respect to the standard analysis.
  Theoretical uncertainties are given by the matching scheme ambiguity,
  the choice of $\xmu$ and the uncertainty 
  of the Milan factor ${\cal M}$. 
  Experimental uncertainties are taken into account by varying the
  input data of the fit.
  All errors  are treated as asymmetric uncertainties on \azero.
}
\end{table}
%
% Table of chis^2/dof results
%
\newpage
\begin{table}[!htb]
\begin{center}
\begin{tabular}{|r||c|c|c|c|}%|c|}
\hline
                   &$\mathbf{1-T}$        &$\mathbf{C}$        &$\mathbf{B_T}$        &$\mathbf{B_W}$\\
\hline
ln(R)     &  283.7/277  &   162.9/170 &   161.0/171 &   162.1/132  \\
\it corr. & \it  -76\%  & \it   -60\% &  \it  -77\% & \it -57\%    \\
\hline 
ln(R) mod.&  341.8/277  &   185.9/170 &   158.2/171 &   133.0/132  \\
\it corr. & \it  -78\%  & \it   -72\% &  \it  -84\% & \it -63\%    \\
\hline
R         &  269.7/277  &   155.9/163 &   179.5/171 &   177.1/132  \\
\it corr. & \it  -75\%  & \it   -61\% &  \it  -76\% & \it -56\%    \\
\hline
R mod.    &  329.0/277  &   171.4/163 &   160.7/171 &   139.3/132  \\
\it corr. & \it  -78\%  & \it   -75\% &  \it  -84\% & \it -64\%    \\
\hline%.
\end{tabular}
\end{center}
\caption{\label{tab-chisq} 
Values for \chisqd\ from the (\as,\azero)-fits of combined power corrections plus
resummed QCD predictions using the for different matching schemes.
Also shown are the correlation coefficients of the fits.
In the case  of some $C$-parameter distributions, the fit ranges 
for the R- and the modified R-matching 
had to be chosen slighty different from those of the
ln(R)- and the modified ln(R) matching 
in order to ensure the convergence of the fit.
%cut by one bin at the lower edge
}
\end{table}

%%%%%%%%%%%%%%%%%%%%%%%%%%%%%%%%%%%%%%%%%%%%%%%%%%%%%%%%%%
%  The figures:
\clearpage
\section*{ Figures }
%
% Data and fit curves for thrust
%

%
% Data and fit curves for C
%
\newpage
%
% Data and fit curves for BT and BW
%
\newpage

%
% Plot of alpha_0 results 
%
\newpage

%%%%%%%%%%%% The End

\end{document}